\documentclass{jfm}
\usepackage{graphicx,subfigure}
\usepackage{epstopdf, epsfig}
\usepackage[abs]{overpic}
\usepackage{xcolor}

\shorttitle{Wave generation by a granular collapse}
\shortauthor{M. Robbe-Saule \emph{et al.}}

\title{Experimental investigation of tsunami waves generated by granular collapse into water}

\author{Manon~Robbe-Saule\aff{1}, Cyprien~Morize\aff{1} \corresp{\email{cyprien.morize@universite-paris-saclay.fr}}, Robin~Henaff\aff{1},
    Yann~Bertho\aff{1}, Alban~Sauret\aff{2}, \and Philippe~Gondret\aff{1}}

\affiliation{\aff{1}Universit\'e Paris-Saclay, CNRS, Laboratoire FAST, 91405, Orsay, France. \aff{2}Department of Mechanical Engineering, University of California, Santa Barbara, CA 93106, USA.}

\begin{document}

\maketitle

\begin{abstract}
The generation of a tsunami wave by an aerial landslide is investigated through model laboratory experiments. We examine the collapse of an initially dry column of grains into a shallow water layer and the subsequent generation of waves. The experiments show that the collective entry of the granular material into water governs the wave generation process. We observe that the amplitude of the wave relative to the water height scales linearly with the Froude number based on the horizontal velocity of the moving granular front relative to the wave velocity. For all the different parameters considered here, the aspect ratio and the volume of the column, the diameter and density of the grains, and the height of the water, the granular collapse acts like a moving piston displacing the water. We also highlight that the density of the falling grains has a negligible influence on the wave amplitude, which suggests that the volume of grains entering the water is the relevant parameter in the wave generation.
\end{abstract}

\section{Introduction}
Tsunami waves are one of the natural phenomena that lead to the most catastrophic disasters on Earth as demonstrated by recent events in Indonesia in 2004 \citep{Titov2005} and in Japan in 2011 \citep{Mori2011}. These nonlinear waves of large wavelength propagate over long distances at high velocity and can reach large amplitudes when arriving on coastlines, which leads to considerable inland damage. To establish hazard assessment maps, it is crucial to evaluate the run-up of such waves over a coast. These dynamics have been the topic of many studies, such as the experiments of \citet{Jensen2003}, the theoretical analysis of \citet{Carrier2003} and \citet{Madsen2010} or the numerical simulations of \citet{Liu2005}. Although the propagation of tsunami waves and their run-up when they arrive on the coast are now reasonably well described, a critical issue is the generation of such waves.

Tsunami waves can be generated from different mechanisms: tectonic or volcanic events, landslides and even impact of meteorites or atmospheric events. Earthquakes are the most important source of tsunamis and were the cause of the two dramatic events in 2004 and 2011. In such cases, the tsunami waves are triggered by the sudden displacement of the seabed during earthquakes \citep{Jamin2015}. The generation of tsunami waves by the impact of meteorites is rare \citep{Bryant2014}, but it may explain some deposits that can be observed on the surface of Mars as claimed recently by \citet{Costard2017}. Considering the recent enhanced activity in the study of tsunami waves and the mechanisms underlying their generation, \citet{Monserrat2006} considered ``tsunami-like" waves that are induced by atmospheric processes, such as atmospheric gravity waves, pressure jumps, frontal passages, squalls and other types of atmospheric disturbances. These high-energy ocean waves occur only for very specific combinations of resonant effects. The rareness of such combinations is the main reason why major ``meteotsunamis'' are exceptional and observed only at a limited number of sites in the world's oceans \citep{Monserrat2006}. However, these exceptional meteorological events may explain some tsunami-like events of ``unknown origin''. Tsunamis generated by landslides are not rare as demonstrated by the recent event, reported by \citet{Grilli2019} and \citet{Paris2020}, which occurred in December 2018 from the partial flank collapse of Anak Krakatau, a few years after being predicted by \citet{Giachetti2012}. Many volcanic islands are prone to such a collapse with an associated risk of tsunamis such as La R\'eunion in the Indian Ocean \citep{Kelfoun2010} and La Palma in the Atlantic Ocean where a giant event may occur and generate a massive wave of hundreds of meters high \citep{Abadie2012}. Such tsunami waves are generated by the displacement of water from subaerial or submarine landslides, which may be triggered by volcanic or seismic events. Landslides may occur in oceans but also in lakes or rivers \citep{Kremer2012, Couston2015}, with the collapse or avalanche of either soil, rocks or snow \citep{Zitti2016}.

In the tsunami generation by aerial landslides, that is, by the entry of some materials into water, the Froude number, corresponding to the ratio of the landslide velocity to the water wave velocity, is expected to play a crucial role. The wave generation caused by a falling mass has been studied first in the early work of John Scott Russel who discovered the solitary wave and then attempted to reproduce it by dropping a solid mass vertically with respect to the water surface, resulting in a constant Froude number of order one \citep{Russell1844}. Recent experiments of \citet{Monaghan2000} using the Scott Russel's wave generator, together with numerical simulations, lead to a nonlinear scaling law for the wave amplitude with the falling mass, which includes a power-law exponent of 2/3. A laboratory tsunami wave can also be generated by the horizontal motion of a vertical plate at a given velocity. In this situation, the amplitude of the wave scales about linearly with the Froude number as observed experimentally by \citet{Miller1966} and \citet{Das1972}, and predicted theoretically by \citet{Noda1970}. More sophisticated laboratory experiments have then been developed to generate tsunami-like waves from model aerial landslides. For instance, \citet{Heinrich1992} considered the impact of a sliding wedge on an inclined plane. More recently, using a similar set-up, \citet{Viroulet2013a} observed different scalings for Froude numbers smaller or larger than 1/2. To account for the complex granular nature of landslides, experiments performed with granular materials have been considered. By studying the entry of granular materials into water at high velocity from a pneumatically launched box along a smooth inclined plane, \citet{Fritz2003a,Fritz2003b,Fritz2004}, \citet{Heller2010} and \citet{McFall2016} have been able to reach high Froude numbers in the range 1--6. They observed that the wave amplitude scales mainly with the Froude number (with a power exponent of about one), but also depends on other parameters such as the thickness of the slide, the slope angle and the launched mass. More recently, \citet{Viroulet2013b, Viroulet2014} considered the impulse wave generated by the gravity-driven collapse of a granular mass above the water surface down a rough inclined plane. The wave amplitude was found to scale almost linearly with the falling mass, with a power exponent 0.9, and was found to depend on the slope angle.

In this paper, we consider the gravity-driven collapse of an initially dry granular column into water. The collapse of a column of dry grains has been studied extensively for its relevance in geophysical applications. Experiments in a two-dimensional configuration of a rectangular channel done by \citet{Balmforth2005} and \citet{Lajeunesse2005} show that the final deposit is mainly controlled by the aspect ratio $a= H_0/L_0$ of the initial column of height $H_0$ and length $L_0$. The run-out length $\Delta L = L_f - L_0$, corresponding to the lateral extension $L_f$ of the final deposit relative to the initial column, increases with the aspect ratio $a$ with two different regimes $\Delta L/L_0 \sim a$ for $a\lesssim 3$ and $\Delta L/L_0 \sim a^{2/3}$ for $a \gtrsim 3$. These findings have been confirmed by numerical simulations, from either discrete element methods \citep{Zenit2005, Staron2005, Lacaze2008} or continuous modeling with viscoplastic rheology \citep{Lagree2011, Ionescu2015}. The typical time scale for the granular collapse corresponds to the free-fall time scale $\sqrt{2H_0/g}$ based on the initial height of the column $H_0$. For tall grain columns ($a\gtrsim 3$), \citet{Larrieu2006} have shown that the dynamics can be understood from depth-averaged shallow-water equations applied to a thin horizontally spreading layer (with basal Coulomb friction), which is fed with vertical ``raining'' grains (with no horizontal momentum) from the initial column in free fall. The influence of an ambient fluid on the granular collapse has then been investigated experimentally by \citet{Meruane2010}, \citet{Rondon2011}, \citet{Bougouin2017} and \citet{Bougouin2018}, and numerically by \citet{Topin2012}. For immersed cases, the Stokes number that measures the importance of grain inertia relative to viscous forces and the solid/fluid density ratio are important dimensionless numbers that control the avalanche regime \citep{Courrech2003a}. \citet{Zheng2018} reported recently on the release of an initial column of buoyant particles below a free surface so that the spreading takes place at the top free surface. Their experiments exhibit different scaling laws for the run-out length and the characteristic time of the spreading than the usual granular collapse.

In the present paper, the wave generated by the collapse of a column of initially dry grains into water is investigated experimentally. In \S2, we describe the experimental set-up and the phenomenology associated with the generation of the waves induced by the granular collapse. The experimental results are then presented and analysed in \S3. More specifically, we investigate the influence of the column geometry (height, width and aspect ratio), of the grains (diameter and density) and of the water height on the generated wave (amplitude and wavelength). We then focus on the motion of the granular front at the water surface. We demonstrate that the local Froude number based on the ratio of the velocity of the granular front to the velocity of the waves governs the generation of the wave and provide a scaling law for the maximum amplitude of the generated wave relative to the water depth.

\section{Description of the experiments}

\subsection{Experimental set-up}

\begin{figure}
  \centering
  \includegraphics[width=\hsize]{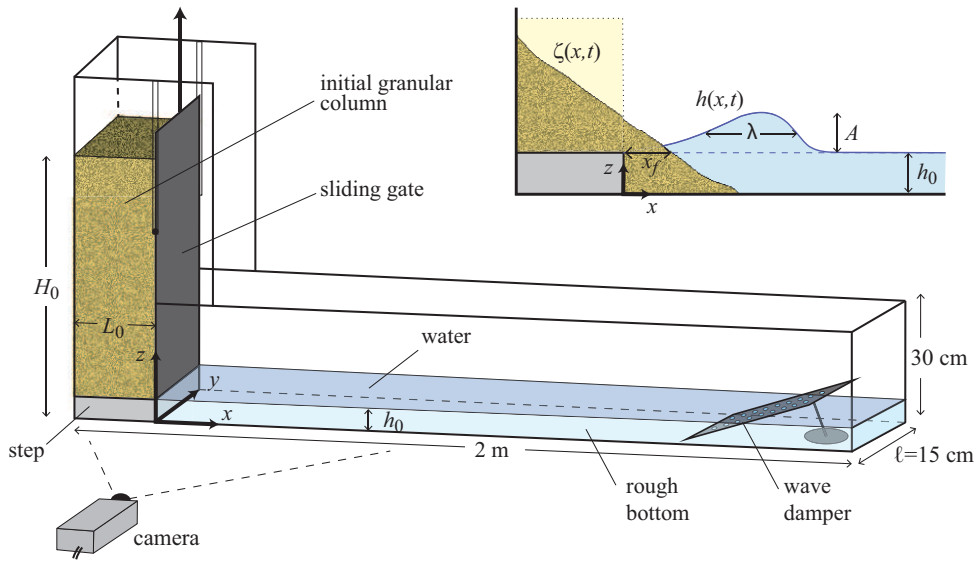}
  \caption{Sketch of the experimental set-up and notation introduced for the collapse of an initial granular column of height $H_0$ and length $L_0$ in a water layer of height $h_0$ within a channel of width $\ell$. The inset shows the notation used to describe the amplitude Here $A$ and width $\lambda$ of the impulse wave, and the granular front $x_f$ at the water surface. $A$ and $\lambda$ are extracted from the position $h(x,t)$ of the water surface: $A$ is the maximal $h(x,t)$ value relative to its initial value $h_0$ and $\lambda$ is the $x$-extension of the wave at mid-height, $\lambda(t) = x_2 - x_1$, where $h(x_1,t )= h(x_2,t) = h_0+A/2$. The position of the granular front $x_f$ is extracted from the position $\zeta(x,t)$ of granular surface at the initial water height $h_0$: $\zeta(x_f,t) = h_0$.}
\label{Fig1}
\end{figure}

The experimental set-up is shown in figure~\ref{Fig1}. It consists of a horizontal rectangular glass tank that is 2~m long, 30~cm high and $\ell =15$~cm wide, which is filled with water over a height $h_0$. A vertical sliding gate maintains initially a column of height $H_0$ and length $L_0$ of monodisperse dry grains of total mass $M$ at one side of the tank (left in figure~\ref{Fig1}). In the following, $a=H_0/L_0$ and $V=H_0L_0\ell$ denote the aspect ratio and the volume of the initial column of grains, respectively. The location of the gate is fixed, but the width $L_0$ of the granular column can be varied by inserting wedges of different sizes along the vertical left-hand sidewall. Upon opening the sliding gate, the subaerial granular column collapses and impacts the water-free surface, generating a wave. Given the aspect ratio of the tank, no significant motion of the grains and fluid takes place in the transverse $y$-direction, which ensures that the system remains quasi-two-dimensional. Glass beads of diameter 3~mm are glued on the channel base to prevent the grains from sliding over the horizontal surface and ensure a no-slip boundary condition. A weight is attached to the gate by a system of ropes and pulleys to reduce the apparent weight of the door, which is quickly removed manually to release the grains that spread in the $x$-direction. The origin $x =0$ of the $x$-axis corresponds to the location of the vertical gate, and the time origin $t = 0$ corresponds to the start of the gate lifting. We ensure that the time required to lift the gate is small compared with the time scale of the granular collapse, so that the opening of the door does not influence significantly the granular collapse \citep{Meriaux2006, Ionescu2015} and, thus, the wave generation.

As a moving wall partially immersed in a fluid would disturb the free surface, the granular column sits on a step of height $h_0$ so that the grains and the door are flush with the surface of the undisturbed water. This ensures that the experiment begins with a completely dry granular medium and is not partially immersed. It was checked that for sufficiently high columns ($H_0/h_0\geq 4$) the collapse of the grains is not affected by the presence of the step, that is, the final deposit height, the run-out length and the dynamics of the collapse remain unchanged with or without step, so that the step can be seen as motionless grains during the collapse. All the experiments presented in this paper satisfy this criterion.

\begin{table*}
\begin{tabular}{p{13mm}p{14mm}p{14mm}p{14mm}p{16mm}p{14mm}p{14mm}p{19mm}}
\hline
Symbol &  $H_0$ (cm) & $L_0$ (cm) & $a$ & $V$ (dm$^3$) & $h_0$ (cm) & $d$ (mm) & $\rho_g$ (kg m$^{-3}$) \\
\\
\Large$\circ$ & [21--42] & [8.5--16.5] & 2.5 & [2.7--10.2] & 5 & 5 & 2500 \\
$\square$ & [20--50] & [5.5--13.5] & [1.5 9] & 4.1 & 5 & 5 & 2500 \\
$\bigtriangleup$ & 35 & [5--20] & [1.8--7] & [2.6--10.2] & 5 & 5 & 2500 \\
$+$ & [20--40] & 10 & [2--4] & [3--6] & 5 & 5 & 2500 \\
$\bigstar$ & 41 & 10 & 4.1 & 6.15 & [2--12] & 5 & 2500 \\
$\times$ & 25 & 10 & 2.5 & 3.75 & 5 & [1--8] & 2500 \\
\Large$\diamond$ & 25 & 10 & 2.5 & 3.75 & 5 & 5 & [1030--7780] \\
\hline
\end{tabular}
\caption{Summary of the different sets of experimental parameters.}
\label{Table1}
\end{table*}

In these experiments, we focus on the wave generation in water of density $\rho_w = 10^3$~kg~m$^{-3}$, viscosity $\eta = 10^{-3}$ Pa s and surface tension $\gamma \simeq 6 \times 10^{-2}$~N m$^{-1}$. The capillary length is $l_c = (\gamma/\rho_w g)^{1/2} \simeq 2.5$ mm. The water height in our experiment is mainly $h_0 = 5$~cm, but we perform additional experiments varying the water height in the range $2 \leq h_0\leq 12$~cm to study its influence on the generated wave. A wave damper, which consists of a perforated Plexiglass plate, is located at the end of the tank to absorb the incoming wave and to avoid reflected waves. The grains used in our experiments are mainly glass beads with a density $\rho_g=2500$~kg~m$^{-3}$ and a diameter $d=5$~mm. We perform additional experiments by varying the glass beads diameter from 1 to 8~mm to characterize the influence of the grain size on the wave generation. We also investigate the influence of the grain density from 1030 to 7780~kg~m$^{-3}$ using light or dense plastic beads, as well as steel beads. For all the grains considered here, the packing fraction $\phi = M/(\rho_gV)$ of the initial column is $\phi \simeq 0.64$, which is the typical value of a dense packing. For all the grain diameters considered here, the number of grains $\ell/d$ in the transverse $y$-direction is at least about 20 so that the lateral wall effects can be considered as negligible in the present unsteady avalanche flow dynamics \citep{Courrech2003b}. The grain/fluid density ratio is in the range $8 \times 10^2 \lesssim \rho_g/\rho_a \lesssim 6 \times 10^3$ for the air and $1.03 \lesssim \rho_g/\rho_w \lesssim 7.78$ for the water. The Stokes number $\mathrm{St} = (\rho_g \Delta \rho g d^3)^{1/2}/(18\sqrt{2}\eta)$ is in the range $5 \times 10^2 \lesssim \mathrm{St} \lesssim 10^4$ for the air and $6 \lesssim \mathrm{St} \lesssim 150$ for the water, where $g$ is the gravitational acceleration and $\eta$ is the dynamic viscosity of the fluid considered. As introduced by \citet{Courrech2003a} for granular avalanches and then used, for instance, by \citet{Rondon2011} and \citet{Bougouin2018}, the fall of grains in the present experiments corresponds to the free-fall regime in the air as $\rho_g/\rho_a > 16$ and $\mathrm{St} > 10$ and to the inertial regime in the water as $\rho_g/\rho_w<16$ and $\mathrm{St}>4$.

The dynamics of the granular collapse and the wave generation, illuminated by a continuous halogen lamp, are imaged through the transparent sidewall of the tank with a video camera with a resolution 1920 $\times$ 1080 pixels and frequency 25~Hz, located 2.5~m in front of the channel. The spatial resolution in these conditions is about 0.9~mm per pixel. Image processing is performed using a custom Matlab routine based on a thresholding method that allows both the instantaneous contour $\zeta(x,t)$ of the granular collapse and the water interface $h(x,t)$ during the wave generation to be determined. Rhodamine dye was added into the water to facilitate the extraction of these quantities by thresholding methods. This procedure allows us to extract both the instantaneous amplitude $A(t)=h(x, t)-h_0$ and the mid-height width $\lambda(t)$ of the impulse wave, and the granular front $x_f(t)$ at the water level as sketched in the inset of figure 1. We performed around 10 runs for the same experimental configuration, and the results show a rather small dispersion of about 10$\%$ around the average value for the different measured quantities. More details can be found in \citet{RobbeSaule2019}.

\begin{figure}
  \centering
  \includegraphics[width=\hsize]{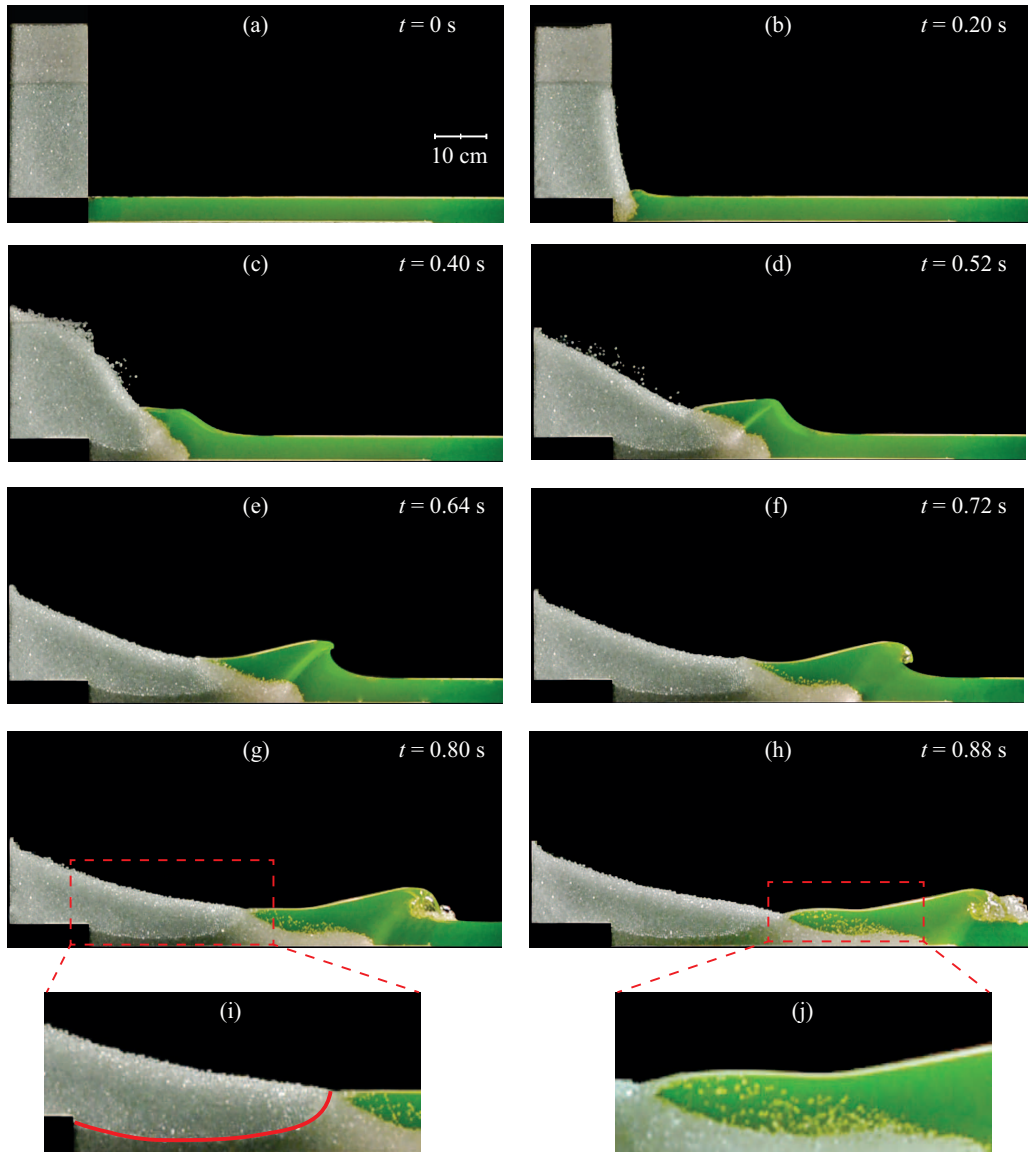}
  \caption{Image sequence showing the generation of a wave by the collapse of a granular column of initial aspect ratio $a=2.5$ ($H_0=41$~cm and $L_0=16.5$~cm) and volume $V=10.2$~dm$^3$ of 5~mm glass beads into $h_0=5$~cm of water; (i) and (j) correspond to enlarged views of (g) and (h), respectively. The red line in (i) indicates the imbibition front separating the dry from the wet granular media. The time $t=0$ is defined as the time from which the door starts to rise.}
\label{Fig2}
\end{figure}

The granular column is expected to collapse with the typical free-fall timescale $\sqrt{2 H_0/g}$ and the typical vertical free-fall velocity $\sqrt{2gH_0}$, whereas the generated water wave is expected to propagate at the typical velocity $\sqrt{gh_0}$ corresponding to gravity waves ($\lambda > 2 \pi l_c$) in shallow water ($\lambda > h_0$). A Froude number corresponding to the ratio of the typical granular velocity to the wave velocity can be defined here as
\begin{equation}
\mathrm{Fr_0} = \sqrt{\frac{H_0}{h_0}},
\end{equation}
which thus depends only on the granular column/water height ratio $H_0/h_0$. For the present experiments with a fixed immersed step at the bottom of the column, $H_0 > h_0$ so that $\mathrm{Fr_0} > 1$.

The collapse dynamics and the corresponding wave generation have been characterized when varying the aspect ratio $a$ and the volume $V$ of the initial granular column, the diameter $d$ and the density $\rho_g$ of the grains as well as the water height $h_0$, as summarized in table~\ref{Table1}.

\subsection{Wave generation from granular collapse}
\label{sec:description}

\begin{figure}
  \centering
  \includegraphics[width=\hsize]{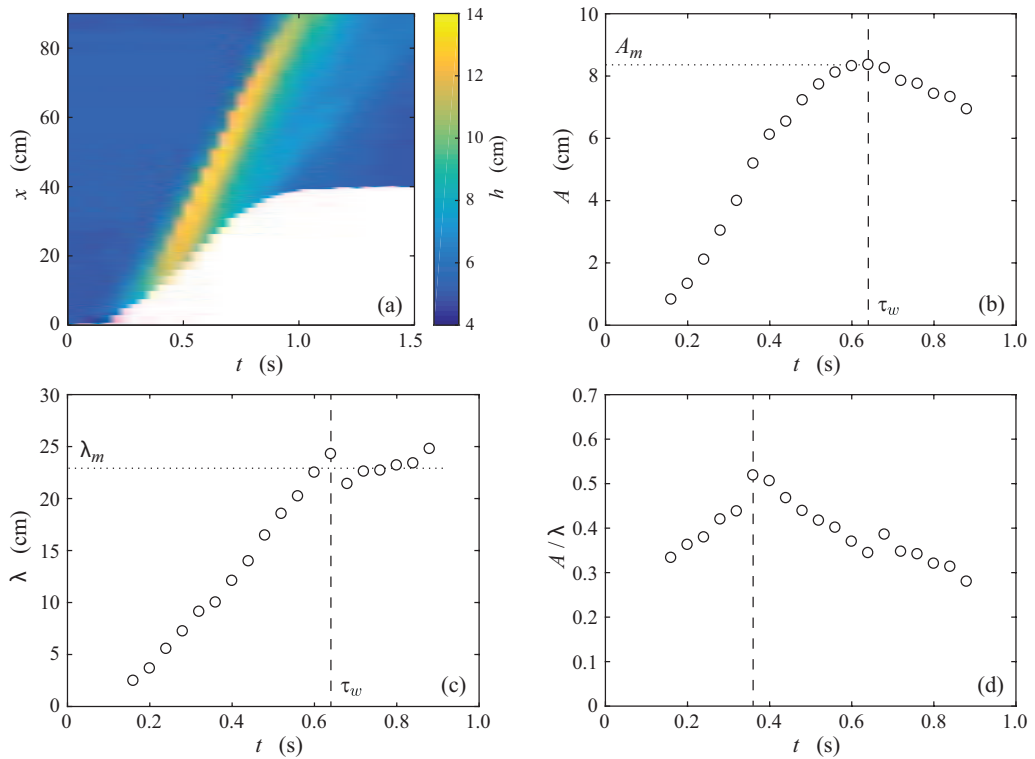}
  \caption{(a) Spatiotemporal evolution of the free-surface elevation $h(x,t)$ for the experiment presented in figure~\ref{Fig2}. The white region corresponds to the spreading of the granular collapse. Time evolution of (b)~the amplitude $A$, (c)~the mid-height width $\lambda$ and (d) the aspect ratio $A/\lambda$ of the leading wave for the experiment reported in figure~\ref{Fig2}. The vertical dashed lines indicate (b,c) the characteristic time $\tau_w \simeq 0.64$~s, where the wave amplitude $A$ reaches its maximum value $A_m$ with the corresponding width $\lambda_m$, and (d) the time $t \simeq 0.36$ s where the aspect ratio $A/\lambda$ is maximum.}
\label{Fig3}
\end{figure}

A typical experiment is shown in figure~\ref{Fig2} where a time lapse illustrates the evolution of the granular collapse and the corresponding generated wave for an initial column of volume $V=10.2$~dm$^3$ and aspect ratio $a=2.5$ of 5~mm glass beads into $h_0=5$~cm of water. The Froude number is $\mathrm{Fr_0}=\sqrt{H_0/h_0} \simeq 2.9$ in this case. The impulse wave is generated when the grains impact the free surface of the water. We observe that the wave amplitude, as well as its width, initially increase [figures~\ref{Fig2}(b)---\ref{Fig2}(d)]. The wave becomes progressively steeper until the wavefront becomes vertical and a jet forms toward the crest [figure~\ref{Fig2}(e)]. The jet splashes down at the surface under the influence of gravity, causing a breaking wave as shown in figures~\ref{Fig2}(f)--\ref{Fig2}(h). The time lapse highlights that the entry of the granular material into water is a complex process involving three different phases: the water, the grains and the air trapped in between grains. Indeed, the penetration of the water in the dry grains is not instantaneous. We observe a well-defined front delimiting dry beads and wet beads appearing at short times below the initial water free surface, as illustrated in figure~\ref{Fig2}(i). This dry/wet front rises progressively during the collapse associated with the imbibition of an equivalent porous medium. As a result, a volume of air is trapped between the grains in the early stage of the collapse and contributes to displacing a given volume of water, and thus contributes to the generation of the wave. Finally, the air trapped in the granular medium escapes, and bubbles rise to the surface as observed in figure~\ref{Fig2}(j).

Figure~\ref{Fig3}(a) shows the spatiotemporal evolution of the water height $h(x,t)$ associated with the experiment of figure~\ref{Fig2}. We see that the leading wave [yellow stripe in figure 3(a)] generated close to the granular front separates from it at $t \simeq 0.5$~s before the end of granular collapse at $t \simeq 1$~s, and then propagates at a constant velocity of about 1.15 m s$^{-1}$ corresponding to the slope of the yellow stripe in figure 3(a). This velocity is significantly larger than the velocity of gravity waves in shallow water $\sqrt{gh_0} \simeq 0.7$~m s$^{-1}$ because the typical wave amplitude $A_m$ is not small when compared with the water depth $h_0$. In our case, the wave velocity is closer to the empirical velocity $\sqrt{g(h_0+A_m)} \simeq 1.14$ m s$^{-1}$ corresponding to a large solitary wave \citep{Russell1844, Boussinesq1872}. Indeed, larger-amplitude waves propagate faster than smaller ones owing to nonlinear effects \citep{Lamb1932}. In addition, the maximum amplitude of the free surface does not occur at the location of the front collapse. We also note the presence of a much weaker secondary wave of millimetric amplitude that propagates much slower than the primary leading wave [thin light blue stripe in figure 3(a)]. This tiny secondary wave can be seen in figure~\ref{Fig2}(h).

\begin{figure}
  \centering
  \includegraphics[width=\hsize]{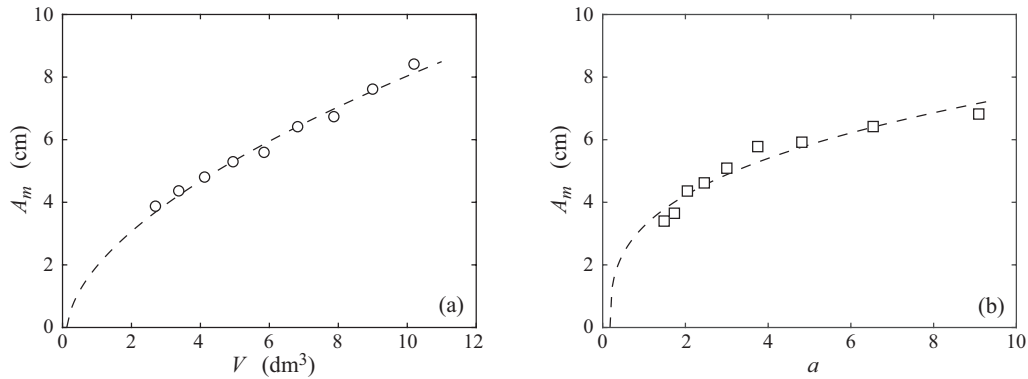}
  \caption{Maximum amplitude $A_m$ of the leading wave when varying (a)~the volume $V$ of the initial column at constant aspect ratio $a = 2.5$ and (b)~the initial aspect ratio $a$ at constant volume $V = 4.1$~dm$^3$: ($\circ$,$\square$) experimental data and (- - -) power law fits (a) $A_m = 2.15 \, (V-0.15)^{0.58}$ and (b) $A_m = 3.9 \, (a-0.09)^{0.34}$.}
\label{Fig4}
\end{figure}

The time evolution of the amplitude $A$, the mid-height width $\lambda$ and the aspect ratio $A/\lambda$ of the leading wave are reported in figures~\ref{Fig3}(b), 3(c) and 3(d), respectively. The wave amplitude and the mid-height width first increase with time. At the characteristic time $\tau_w \simeq 0.64$~s the wave reaches its maximum amplitude value $A_{m} \simeq 8.3$~cm with the corresponding width $\lambda_m \simeq 23$~cm. The amplitude of the wave then decreases whereas the width of the wave remains approximately constant. Looking now at the aspect ratio of the wave [figure 3(d)], we observe that it first increases, which means that the amplitude increases first faster than the width, and then decreases after $t \simeq 0.36$ s. This time for which the aspect ratio reaches the maximum value $A/\lambda \simeq 0.5$ corresponds to an inflection point in the evolution of $A(t)$ whereas the time evolution of $\lambda(t)$ is not modified. In all of the experiments listed in table~\ref{Table1}, the wave breaks as soon as the slope $| \partial h/\partial x|$ of its upstream face is about $0.34 \pm 0.03$, in good agreement with the value 0.32 reported by \citet{Deike2015} at large Bond number, that is, when gravitational forces are large compared with surface tension forces. Beyond $t\simeq 0.9$~s, the wave gradually exits the camera field of view and we stop the measurements record.

In the following section, we first present the results on the leading wave generated by the collapse of the granular column when varying the different parameters of the set-up and then focus on the dynamics of the granular front.

\section{Experimental results}

\subsection{Influence of the geometry of the initial column}

In our configuration, both the aspect ratio of the initial column of grains and the volume are expected to play an important role in the wave generation. Figure 4 displays the results from experiments made either by varying the volume of the column in the range $2.7 \lesssim V \lesssim 10.2$~dm$^3$ at a constant aspect ratio $a=2.5$, or by varying the aspect ratio in the range $1.5 \lesssim a \lesssim 9$ at a constant volume $V = 4.1$~dm$^3$, to characterize the influence of $V$ and $a$, respectively. We observe that the maximum amplitude $A_m$ of the leading wave increases both with the volume $V$ and with the aspect ratio $a$ of the initial column. The increase of $A_m$ with $V$ is not quite linear as shown in figure 4(a). In our experimental configuration with a fixed bottom part of the column of height $h_0$, the amplitude of the generated wave is expected to be zero for $H_0 = h_0$ corresponding here to a volume $V_* = h_0^2 \ell/a = 0.15$ dm$^3$. A power law $A_m \sim (V-V_*)^{\alpha}$ fits well the data with a power exponent $\alpha \simeq 0.58$, significantly lower than one. The increase of $A_m$ with $a$ shown in figure 4(b) is even less linear. The amplitude of the generated wave is here expected to be zero for an aspect ratio $a_* = h_0^2 \ell/V = 0.09$. A power law $A_m \sim (a-a_*)^{\beta}$ also fits quite well the data with a power exponent $\beta \simeq 0.34$ much lower than one. These results show that both the volume and the aspect ratio of the column are important parameters in the wave generation.

\begin{figure}
  \centering
  \includegraphics[width=\hsize]{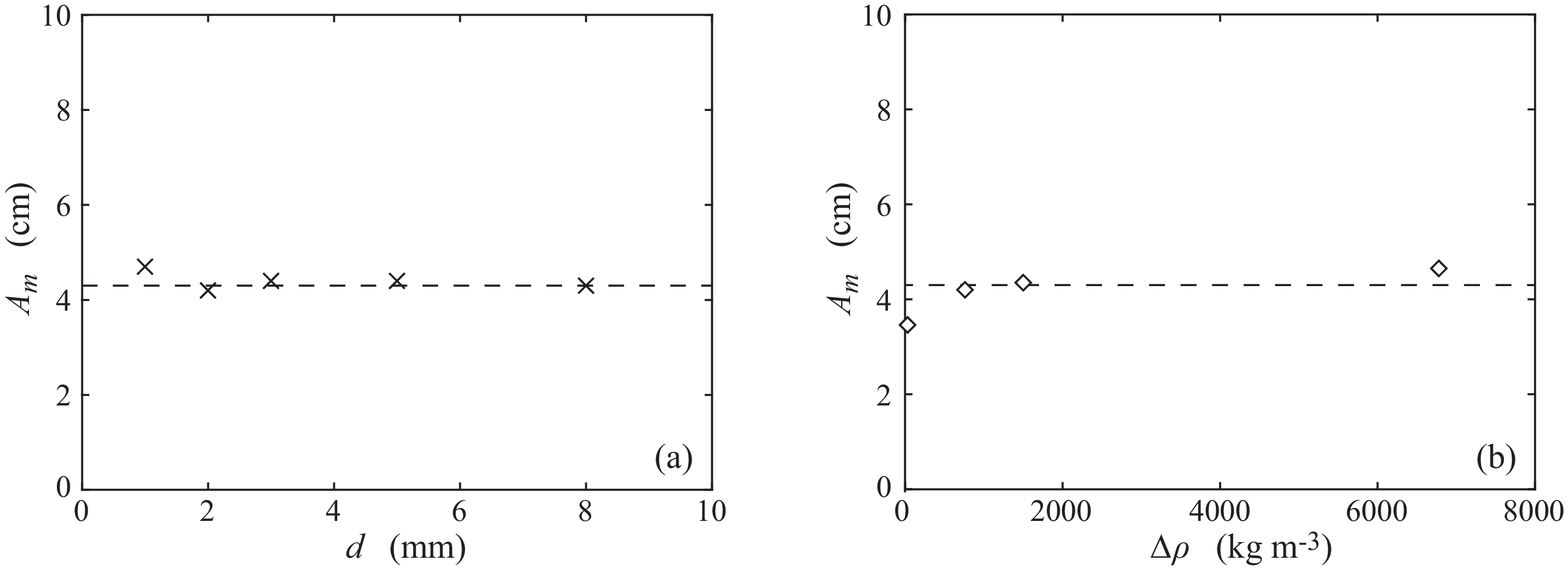}
  \caption{Maximum amplitude $A_m$ of the leading wave as a function of (a)~the grain diameter $d$ and (b)~the difference of density between grains and water $\Delta\rho =\rho_g - \rho_w$, for the collapse of an initial column of aspect ratio $a=2.5$ and volume $V=3.7$~dm$^3$: ($\times$,$\diamond$) experimental data and (- - -) constant value $A_m = 4.3$ cm.}
\label{Fig5}
\end{figure}

\subsection{Influence of the grains}

The influence of the diameter $d$ and density $\rho_g$ of the grains on the amplitude of the leading wave is reported in figure 5. Figure~\ref{Fig5}(a) shows that there is no significant influence of the grain diameter, in the range considered here, $1 \lesssim d \lesssim 8$~mm, on the amplitude of the generated wave, which remains approximately constant with $A_m\simeq 4.3$~cm for the same initial column of aspect ratio $a = 2.5$ and volume $V = 3.7$~dm$^3$. This result suggests that the air initially trapped between the grains remains trapped until the wave has moved away, and the time of imbibition remains larger than the collapse timescale regardless of the size of the grains in the range $0.4 \lesssim d/l_c \lesssim 3$, where $l_c \simeq 2.5$ mm is the capillary length of water. Figure~\ref{Fig5}(b) represents the maximum amplitude $A_m$ of the leading wave as a function of the density difference $\Delta \rho = \rho_g - \rho_w$ between the grains and the water using light and dense plastic beads, as well as glass and steel beads, thus varying the density difference in the range $30 \lesssim \Delta \rho \lesssim 6780$ kg m$^{-3}$. By increasing the density difference by a factor of 226, we note that the wave amplitude only increases by a factor of 1.4. A fit of the data by a power law would lead to a very small power exponent of about 0.05. We can thus conclude that there is no significant effect of the density of the grains on the wave amplitude except at vanishing density difference. This result suggests that the mass of the falling column is a less relevant parameter compared with its volume. Thus, all the results of figures 4 and 5 highlight that the relevant initial parameters for the wave generation in this geometry are the volume (not the mass) and the shape (e.g. aspect ratio) of the granular column, but not the characteristics of the individual grains (e.g. size and density) within the range considered in the present experiments.

\subsection{Influence of the water depth}

\begin{figure}
  \centering
  \includegraphics[width=\hsize]{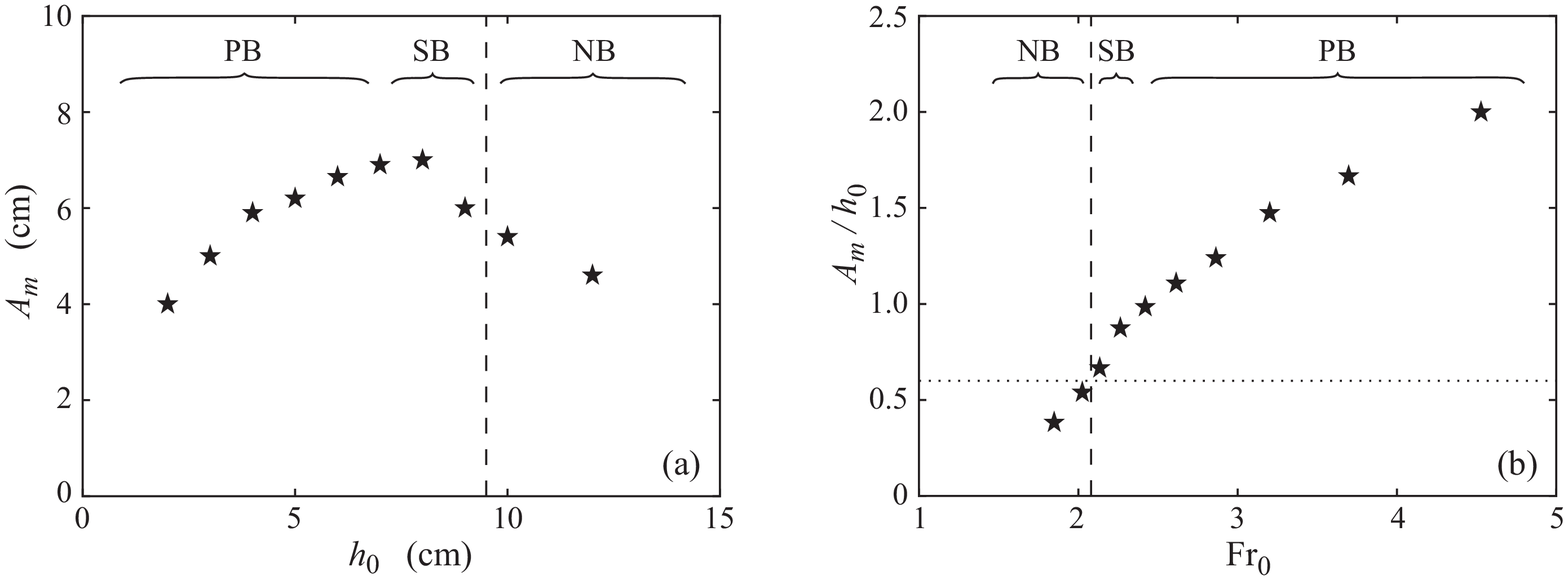}
  \caption{(a)~Maximum amplitude of the leading wave $A_m$ as a function of the initial water height $h_0$ and (b)~relative maximum amplitude of the wave $A_m/h_0$ as a function of the global Froude number $\mathrm{Fr_0} = \sqrt{H_0/h_0}$ for an initial column of volume $V=6.15$~dm$^3$ and aspect ratio $a=4.1$ ($H_0 = 41$ cm). The vertical dashed line indicates the transition observed between non-breaking waves (NB) and breaking waves, the latter corresponding either to plunging breakers (PB) or spilling breakers (SB). The horizontal dotted line corresponds to the criterion $A_m/h_0 \simeq 0.6$ of \citet{Bullard2019} for the transition between breaking and non-breaking waves.}
\label{Fig6}
\end{figure}

We now investigate the influence of the water depth $h_0$ on the maximum amplitude $A_m$ of the leading wave. The results are shown in figure~\ref{Fig6}(a) where the water depth is varied in the range $2 \leqslant h_0 \leqslant 12$~cm for an initial column of constant aspect ratio $a=4.1$ and volume $V=6.15$~dm$^3$. We observe that the maximum amplitude of the wave does not exhibit a monotonic dependence with $h_0$ as it first increases and then decreases. Such an evolution has been recently reported by \citet{Miller2017} and \citet{Bullard2019}. In the present configuration, $A_m$ first increases with $h_0$ up to a maximum [$A_m \simeq 7$ cm for $h_0 \simeq 8$~cm in figure 6(a)] and then decreases for deeper water. Indeed, the wave amplitude is expected to vanish both for $h_0 \to 0$ and for $h_0 \to H_0 = 41$ cm here. Note that the generated waves are of different types when varying the water depth: breaking waves are observed for small enough water depth ($h_0 \lesssim 9.5$~cm), whereas non-breaking waves are observed for large enough water depth ($h_0 \gtrsim 9.5$~cm). Note also that breaking waves can be separated into plunging breakers for $h_0 \lesssim 7$~cm, which correspond to a crest overturning jet, and spilling breakers for $7 \lesssim h_0 \lesssim 9.5$~cm, which are generally defined as waves that collapse without any air entrainment by overturning as described by \citet{Deike2015}. Spilling waves correspond thus to a transitional regime between plunging waves and non-breaking waves.

The data set presented in figure~\ref{Fig6}(a) is reported in figure~\ref{Fig6}(b) in terms of dimensionless quantities: the wave amplitude relative to the water height $A_m/h_0$ is plotted as a function of the global Froude number $\mathrm{Fr_0}=\sqrt{H_0/h_0}$. The ratio $A_m/h_0$ displays a monotonic increase with $\mathrm{Fr_0}$: non-breaking waves correspond to small $\mathrm{Fr_0}$ values whereas breaking waves correspond to large $\mathrm{Fr_0}$ values. The criterion $A_m/h_0\simeq 0.6$ proposed recently by \citet{Bullard2019} for the transition from breaking waves to non-breaking waves, slightly different from the criterion $A_m/h_0\simeq 0.78$ of \citet{McCowan1894}, captures well our experimental observations. In the non-breaking cases corresponding to low enough $\mathrm{Fr_0}$, the shape of the leading wave is well fitted by the symmetric $sech^2$ curve of a theoretical soliton from the Korteweg--de Vries equations \citep{Dauxois2006}. In the breaking cases corresponding to high $\mathrm{Fr_0}$, the wave shape is no more symmetric with a steeper upstream slope.

\subsection{Influence of the global Froude number}

Our experimental results highlight that the generation of a tsunami wave by the collapse of a granular column is a complex system involving many physical parameters. We observe that the amplitude of the generated leading wave does not depend significantly on the properties (size and density) of the grains but depends on the characteristics of the whole grain assembly (volume and aspect ratio of the initial column) and on the water depth.

\begin{figure}
  \centering
  \includegraphics[width=\hsize]{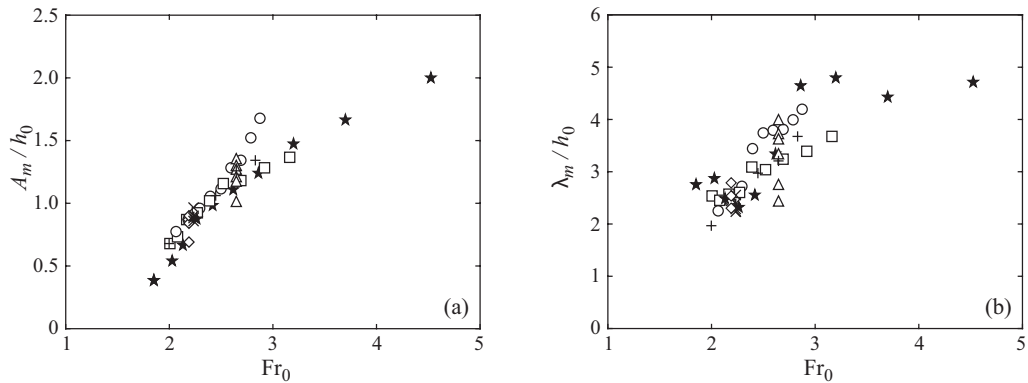}
  \caption{(a) Maximum relative amplitude $A_m/h_0$ and (b) relative width $\lambda_m/h_0$ of the leading wave as a function of the global Froude number $\mathrm{Fr_0}=\sqrt{H_0/h_0}$ for all the experiments listed in table 1.}
\label{Fig7}
\end{figure}

As observed in figure 6(b), the global Froude number $\mathrm{Fr_0}=\sqrt{H_0/h_0}$ is expected to be a key parameter for the wave generation in the present configuration. The evolution of the dimensionless amplitude $A_m/h_0$ and the corresponding dimensionless width $\lambda_m/h_0$ of the leading wave are shown in figures~\ref{Fig7}(a) and \ref{Fig7}(b) as a function of $\mathrm{Fr_0}$ for all the experiments reported in table 1. Despite some dispersion of the data, there is a clear increase of both $A_m/h_0$ and $\lambda_m/h_0$ with $\mathrm{Fr_0}$. However, we see that the global Froude number $\mathrm{Fr_0}$ does not govern alone the wave generation because the experiments performed at constant $\mathrm{Fr_0}$ but at different aspect ratio $a$, by varying the column width $L_0$ (and volume $V$), lead to significantly different values $A_m/h_0$ and $\lambda_m/h_0$ (see triangular symbols ($\triangle$) in figure~\ref{Fig7}).

\subsection{Wave velocity}

\begin{figure}
  \centering
  \includegraphics[width=\hsize]{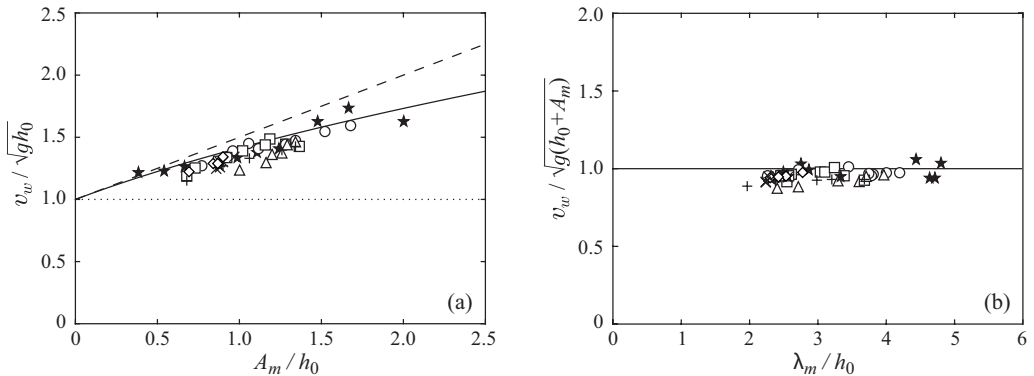}
  \caption{(a) Velocity of the leading wave $v_w$ normalized by the gravity wave velocity $\sqrt{gh_0}$ in shallow water as a function of the relative maximum amplitude of the wave $A_m/h_0$. The symbols corresponds to all the experiments listed in table~\ref{Table1}. The dotted line ($\cdots$) indicates the theoretical value for vanishing amplitude, the dashed line (- - -) corresponds to the theoretical value $1+A_m/2h_0$ from the Korteweg--de Vries equation for solitary waves of small amplitude, and the solid line (---) denotes the empirical value $\sqrt{1+A_m/h_0}$ for solitary waves of large amplitude \citep{Russell1844}. (b)~Wave velocity $v_w$ normalized by Russel's empirical velocity $\sqrt{g(h_0+A_m)}$ as a function of the dimensionless wavelength $\lambda_m/h_0$ for the same data points and with the same empirical value (---).}
  \label{Fig8}
\end{figure}

The impulse wave generated by the collapse of the granular column evolves in time and propagates as shown in the spatiotemporal plot in figure~\ref{Fig3}(a). We have measured the wave velocity $v_w$ by tracking the leading front of the wave as a function of time for every experimental run. The results for the many experiments are shown in figure~\ref{Fig8}(a) where the wave velocity $v_w$, made dimensionless with the velocity $\sqrt{gh_0}$ of gravity waves in shallow water, is plotted as a function of the maximum amplitude of the wave relative to the water depth $A_m/h_0$. All data points correspond to a velocity of the wave larger than $\sqrt{gh_0}$, which corresponds to the limit case of gravity waves of very small amplitude. The data are instead well described by the empirical velocity $\sqrt{g(h_0+A_m)}$ of \citet{Russell1844} for solitary waves of large amplitude, even if this velocity slightly overestimates our measurements. The analytical velocity $\sqrt{gh_0}(1+A_m/2h_0)$ that arises from the non-linear Korteweg--de Vries equation \citep{Dauxois2006} leads to a larger overestimate. Indeed, this velocity is valid for finite but small wave amplitude ($A_m/h_0 \ll 1$), which is not the case in our experiments where $A_m/h_0$ is of order one. The same measurements for the wave velocity now normalized by Russel's velocity $\sqrt{g(h_0+A_m)}$ are plotted in figure 8(b) as a function of the ratio of the wavelength to the water depth $\lambda_m/h_0$. All the data points collapse on the theoretical horizontal line and we observe that they all correspond to shallow water waves with $\lambda_m/h_0 \gtrsim 2$.

\subsection{Energy transfer}

\begin{figure}
  \centering
  \includegraphics[width=\hsize]{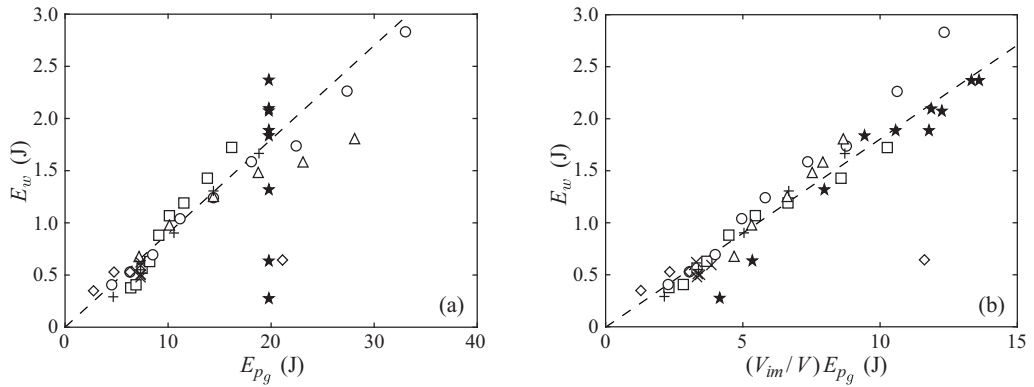}
  \caption{Energy of the generated wave $E_w$ as a function (a) of the potential energy $E_{p_g}$ of the initial granular column and (b) of the fraction of energy $(V_{im}/V)\: E_{p_g} $ of the initial granular column that enters ultimately into water, for all the experiments listed in table 1. The dashed lines correspond to (a) $E_w = 0.09 \: E_{p_g}$ and (b) $E_w = 0.18 \: (V_{im}/V ) \: E_{p_g}$.}
\label{Fig9}
\end{figure}

We now look at an estimate of the transfer of energy from the initial granular column to the water wave. The potential energy $E_{p_g}$ associated with initial granular column is

\begin{equation}
E_{p_g} = M g \frac{H_0}{2} = \frac{1}{2} \phi \rho_g g \ell L_0 H_0^2.
\end{equation}

The energy $E_w$ of the generated wave of volume $V_w$ is the sum of its potential energy $E_{p_w} \simeq \rho_w V_w g A_m/2$ and of its kinetic energy $E_{k_w} = \rho_w V_w v_w^2/2$. Taking $V_w \simeq A_m \lambda_m \ell$ for the volume of the wave and $v_w \simeq \sqrt{g(h_0+A_m)}$ for its velocity as shown in \S3.5, the energy of the generated wave is
\begin{equation}
E_w = E_{p_w} + E_{k_w} \simeq \rho_w g \ell \lambda_m A_m^2 \left(1+\frac{h_0}{2A_m} \right),
\end{equation}

where the term $1+h_0/2A_m$ varies typically between about 1.2 and 2 in our experiments. The energy of the generated wave $E_w$ is plotted as a function of the potential energy of the initial granular column $E_{p_g}$ in figure~\ref{Fig9}(a). The data points corresponding to initial columns of glass beads falling into a water layer of depth $h_0 = 5$ cm collapse roughly on a straight line of slope 0.09 meaning that the corresponding energy transfer is about 9\%. This low ratio is not surprising as it is known that there is a significant dissipation in granular flows. During the collapse of dry granular columns on a flat rough bottom, only about 20\% of the initial potential energy of the column is transferred into kinetic energy of grains \citep{Staron2005}. We also see that the energy transfer from the initial granular column to the water wave strongly depends on the water depth: it varies between about 1\% and 12\% when $h_0$ varies between 2 and 12 cm (see $\bigstar$). This strong effect of the water depth is related to the fact that only a fraction of the grains in motion can transfer their energy to the water since only a fraction of grains falls into water. This fraction of grains corresponds to the ratio of the final volume of grains immersed into water $V_{im}$ relative to the initial volume of the column $V$. The energy of the generated wave as a function of the fraction $(V_{im}/V) \: E_{p_g}$ of potential energy of the initial column of grains that goes into water at the end of the collapse is reported in figure~\ref{Fig9}(b). We see that most of the data points collapse on a straight line of slope $0.18 \pm 0.03$, meaning that the energy transfer from the dry grains falling into water and the generated wave is now about 18\%, close to the fraction of 20\% reported by \citet{Staron2005} for the energy transfer in dry granular collapses. This suggests that the energy losses arise mainly from dissipation in the granular flow itself and not in the coupling of grains with water. This suggestion is related to the fact that the final deposit at the end of the collapse is similar when compared with the dry case, as the grain motion in water corresponds to an inertial regime of high Stokes number ($St > 4$) \citep{Courrech2003a, Topin2012}. Note that the data point corresponding to steel beads is far below the dashed line of slope 0.18. The much weaker energy transfer, of about 5\%, is because the granular collapse for steel and glass beads is similar, so that the energy dissipated in the granular flow is higher for steel than for glass beads.

\begin{figure}
  \centering
  \includegraphics[width=\hsize]{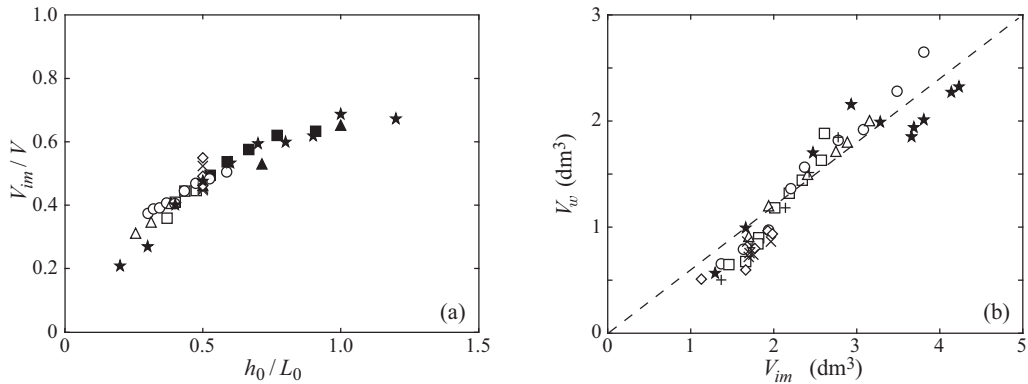}
  \caption{(a) Volume fraction of immersed grains relative to the initial volume of the column, $V_{im}/V$, as a function of the rescaled water depth $h_0/L_0$, for all the experiments listed in table 1 with here filled symbols for $a > 3$. (b)~Volume of the wave $V_w$ as a function of the final volume $V_{im}$ of grains immersed into water for the same data and with a dashed line of equation $V_w \simeq 0.6 V_{im}$.}
\label{Fig10}
\end{figure}

The volume of grains that fell into water is a key parameter for the wave generation. As reported in figure 10(a), we observe that the immersed volume of grains at the end of the collapse, that is, the fraction of grains that fell in water, rescaled by the initial volume of the column is mainly governed by the ratio of the water depth and the initial width of the column. Indeed, figure 10(a) demonstrates that such a rescaling allows collapsing all our experiments on a master curve. This trend can be understood again from the scaling laws reported by \citet{Balmforth2005} and \citet{Lajeunesse2005} for the collapse of a dry granular column of aspect ratio $a = H_0/L_0$ on a horizontal plane, where the run-out length $\Delta L$ is given by $\Delta L/L_0 \sim a$ for $a\lesssim 3$, and $\Delta L/L_0 \sim a^{2/3}$ for $a \gtrsim 3$, which implies $\Delta L \sim H_0$ for $a\lesssim 3$ and $\Delta L \sim H_0^{2/3}L_0^{1/3}$ for $a \gtrsim 3$. In our configuration of granular collapse with a free-fall regime in air ($\rho_g/\rho_a > 16$, $\mathrm{St} > 10$) and inertial regime in water ($\rho_g/\rho_w<16$, $\mathrm{St}>4$) from the regime diagram of \citet{Courrech2003a}, we indeed observe that the run-out and dynamics are similar to the pure dry case. The final volume of grains that fell in water should scale at first order as $V_{im} = \Delta L h_0 \ell$, thus as $V_{im} \sim H_0 h_0 \ell$ for $a\lesssim 3$ and as $V_{im} \sim H_0^{2/3} L_0^{1/3} h_0 \ell$ for $a \gtrsim 3$. This leads to the scalings $V_{im}/V \sim h_0/L_0$ for $a\lesssim 3$ and $V_{im}/V \sim (h_0/L_0)a^{-1/3}$ for $a \gtrsim 3$. Thus, the fraction of immersed grains $V_{im}/V$ is expected to scale approximatively linearly with $h_0/L_0$ with a weak possible effect of the aspect ratio $a$ of the collapsing column (symbols are filled in figure 10(a) for $a > 3$).

Figure~\ref{Fig10}(b) shows that the volume $V_w$ of the generated wave increases with the volume $V_{im}$ of grains that fell into water.
The experimental data can be fitted by the relation $V_w \simeq 0.6 V_{im}$, with a coefficient of proportionality close to the packing fraction $\phi$. The dispersion of the data around this fit is likely induced by some variations caused by the structure of the granular packing. To provide a deeper understanding of the wave generated by the entry of the granular medium into water, we now focus on the dynamics of the collapse at the water surface.

\subsection{Dynamics of the granular front}

\begin{figure}
  \centering
  \includegraphics[width=\hsize]{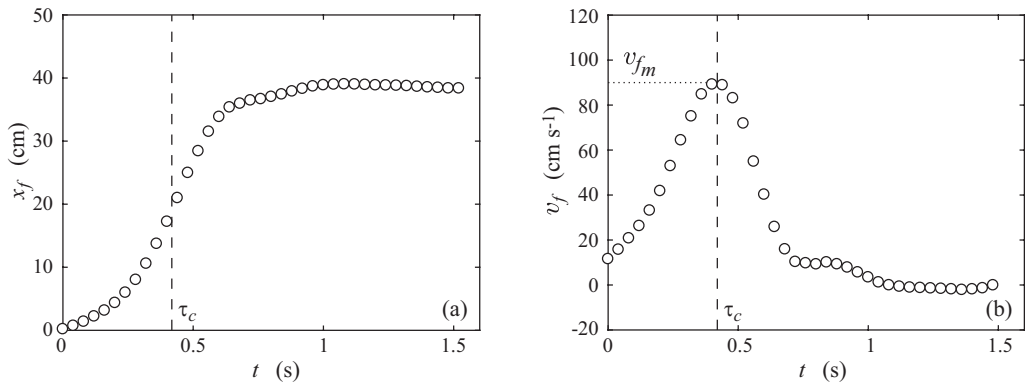}
  \caption{Time evolution of (a)~the horizontal front of the granular collapse $x_f$ at $z=h_0$ and (b)~the horizontal velocity of the granular front $v_f$ at $z=h_0$ for the experiment of figure 2. The vertical dashed line indicates the characteristic time $\tau_c$ where the front velocity is maximum with the value ${v_f}_m$.}
\label{Fig11}
\end{figure}

As the way the grains impact and enter the water is expected to be crucial on the generated wave, we focus here on the position of the instantaneous granular front $x_f(t)$ at the water height $z=h_0$. This granular front corresponds to $\zeta(x_f,t)=h_0$, where $\zeta(x,t)$ depicts the top contour of the grains as shown in the inset of figure 1. Figure~\ref{Fig11}(a) displays an example of the time evolution of the granular front $x_f$ for the experiment reported in figures 2 and 3. The granular front quickly moves forward for approximately 0.65 s and then slows down before coming to rest. The front velocity is defined as $v_f=\mathrm{d}x_f/\mathrm{d}t$, which corresponds to the horizontal velocity of the granular front at the water height $z=h_0$. The time evolution of $v_f$ is shown in figure~\ref{Fig11}(b): $v_f$ quickly increases up to a maximum value $v_{f_m} \simeq 90$~cm s$^{-1}$ at time $\tau_c \simeq 0.4$~s with a typical acceleration 2.3 m s$^{-2}$ lower than the gravitational acceleration. This maximum corresponds to the inflection point of the increase of $x_f(t)$ in figure~\ref{Fig11}(a). Beyond $\tau_c$, the front velocity quickly decreases with a typical deceleration -3 m s$^{-2}$ before vanishing after $t \simeq 0.65$~s $\simeq 1.6 \, \tau_c$. In this example, the characteristic time $\tau_c$ corresponds to $\tau_c \simeq 1.4 \sqrt{2 H_0 /g}$, and the spreading of the front lasts for about $2.2 \sqrt{2 H_0 /g}$. As already discussed in \S2.2, $\tau_c$ corresponds also to the inflection point for the time evolution of the wave amplitude $A(t)$ where the aspect ratio of the wave is maximal [figures 3(b) and 3(d)].

\begin{figure}
  \centering
  \includegraphics[width=0.5\hsize]{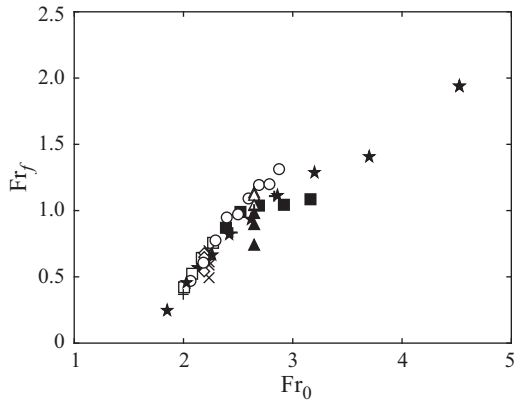}
  \caption{Local Froude number $\mathrm{Fr}_f = {v_f}_m/\sqrt{g h_0}$ as a function of the global Froude number $\mathrm{Fr_0}=\sqrt{H_0/h_0}$ for all the experiments listed in table 1 with filled symbols for $a > 3$.}
\label{Fig12}
\end{figure}

The maximal horizontal velocity of the granular front ${v_f}_m$ depends both on the initial column and on the water height, because the granular front is considered at the water height $h_0$. Using this typical front velocity at the water surface, one can define a local Froude number $\mathrm{Fr}_f$ as
\begin{equation}
\mathrm{Fr}_f = \frac{{v_f}_m}{\sqrt{g h_0}}.
\end{equation}
The local Froude number is based on the horizontal velocity of the granular front whereas the global Froude number $\mathrm{Fr_0}=\sqrt{H_0/h_0}$ is based on the typical (vertical) free-fall velocity of grains. Figure~\ref{Fig12} shows the evolution of the local Froude number $\mathrm{Fr}_f$ as a function of the global Froude number $\mathrm{Fr_0}$. Despite some dispersion in the data, there is a clear increase of $\mathrm{Fr}_f$ with $\mathrm{Fr_0}$, even if there is also an effect of the aspect ratio of the column at the same $\mathrm{Fr_0}$ corresponding to the same column height. This trend can be understood again from the recalled scaling laws of \citet{Balmforth2005} and \citet{Lajeunesse2005} for the run-out length of a dry granular column, and from the typical time scale given by $\tau \sim (H_0/g)^{1/2}$. The front velocity, which is at first order ${v_f}_m \sim \Delta L/\tau$, should thus scale at first order as ${v_f}_m \sim (gH_0)^{1/2}$ for $a\lesssim 3$ and ${v_f}_m \sim (gH_0)^{1/2}a^{-1/3}$ for $a \gtrsim 3$. This explains why the local Froude number $\mathrm{Fr}_f = {v_f}_m/(gh_0)^{1/2} $ is governed mainly by the global Froude number $\mathrm{Fr_0} = (H_0/h_0)^{1/2}$ as shown in figure~\ref{Fig12}, with a possible influence of the aspect ratio $a$ of the collapsing column (symbols are here filled for $a > 3$).

\begin{figure}
  \centering
  \includegraphics[width=\hsize]{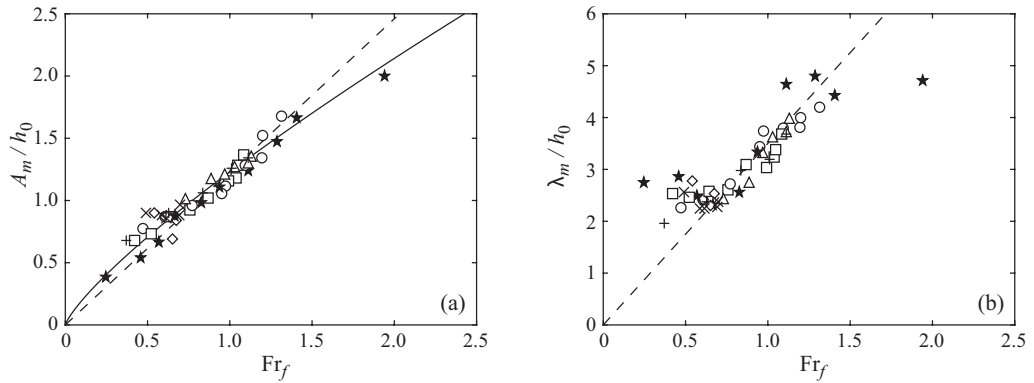}
  \caption{(a) Normalized maximal amplitude $A_m/h_0$ of the wave as a function of the local Froude number $\mathrm{Fr}_f ={v_f}_m/\sqrt{g h_0}$ for all the experiments listed in table~\ref{Table1} (symbols), together with the best linear fit $A_m/h_0=1.23 \, \mathrm{Fr}_f$ (- - -) and the best power law fit $A_m/h_0= 1.23 \, \mathrm{Fr}_f^{0.8}$ (---). (b)~Normalized width $\lambda_m/h_0$ of the wave as a function of $\mathrm{Fr}_f$ with the linear trend $\lambda_m/h_0 \simeq 3.5 \, \mathrm{Fr}_f$ (- - -).}
\label{Fig13}
\end{figure}

The dynamics of the collapse at the water surface control the local Froude number, which may, in turn, govern the generated wave. To check that, the evolutions of the relative maximum amplitude $A_m/h_0$ and the corresponding relative width $\lambda_m/h_0$ of the leading wave are plotted in figures~\ref{Fig13}(a) and \ref{Fig13}(b) as a function of the local Froude number $\mathrm{Fr}_f$. We observe that all the experimental data listed in table~\ref{Table1} exhibit a better collapse on master curves than in figure~\ref{Fig7} where the global Froude number was used. This result suggests that the generated wave is driven by the spreading dynamics of the granular front at the water free-surface that depends on the parameters and the geometry of the initial column of grains. Therefore, in the range of parameters considered here, the wave generation is governed by the collective dynamics of the grains at the interface during the collapse rather than the impact velocity of the individual grains.

The data in figure~\ref{Fig13}(a) are rather well captured by the linear variation $A_m/h_0=1.23~\mathrm{Fr}_f$, with a dispersion of about 15\%. Thus, the waves generated by a granular collapse behave very similarly to the waves produced by a moving piston. Indeed, the nonlinear waves generated by moving horizontally a wall at a constant velocity $V_0$ have been measured experimentally by \citet{Miller1966} and analysed theoretically by \citet{Noda1970}. Their results are close to the linear variation $A_m/h_0 \simeq 1.2 V_0/\sqrt{g h_0}$. As reported by \citet{Das1972}, the linearized wave equations considered by \citet{Noda1970}, which lead to the linear prediction $A_m/h_0 \simeq 1.3 V_0/\sqrt{g h_0}$, have been derived assuming that the acceleration, velocity and displacement of the wall are small compared with $g$, $\sqrt{g h_0}$ and $h_0$, respectively. In our experiments these assumptions hold for the acceleration and velocity of the moving front, but not for its displacement. However, the assumptions for the displacement and the velocity of the piston did not hold in the experiments of \citet{Miller1966} and \citet{Das1972} as well. Similarly to what has been observed by \citet{Das1972} for $A_m/h_0$ as a function of $V_0/\sqrt{gh_0}$ in their experiments with moving piston, our data seem to deviate slightly from the linear law $A_m/h_0 \sim \mathrm{Fr}_f$. A best fit of our data by a power law leads to $A_m/h_0= 1.23 \, \mathrm{Fr}^{0.8}$, with a power exponent slightly smaller than one in our case. The dispersion of the data when considering the best fit by a power law is reduced to about 10\%. Figure~\ref{Fig13}(b) shows that the relative wave width $\lambda_m/h_0$ increases with the local Froude number $\mathrm{Fr}_f$, with the linear trend $\lambda_m/h_0 \simeq 3.5 \, \mathrm{Fr}_f$ except for $\mathrm{Fr}_f \lesssim 0.5$ where it may saturate at $\lambda_m/h_0 \simeq 2.4$, and for $\mathrm{Fr}_f \gtrsim 1.5$ where it may also saturate at $\lambda_m/h_0 \simeq 4.8$.

\begin{figure}
  \centering
  \includegraphics[width=\hsize]{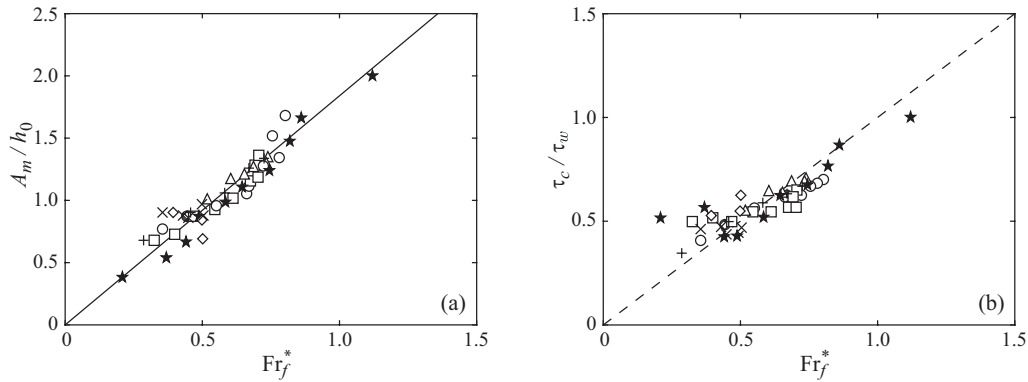}
  \caption{(a) Normalized maximum amplitude $A_m/h_0$ as a function of the modified local Froude number $\mathrm{Fr}_f^* = {v_f}_m/\sqrt{g(h_0+A_m)}$ based on the velocity of solitary waves, for all the experiments listed in table~\ref{Table1} (symbols) together with the best linear fit $A_m/h_0=1.84~\mathrm{Fr}_f^*$ (---). (b) Time ratio $\tau_c/\tau_w$ as a function of $\mathrm{Fr}_f^*$. The dashed line corresponds to $\tau_c/\tau_w = \mathrm{Fr}_f^*$.}
\label{Fig14}
\end{figure}

However, because of the nonlinear nature of the solitary impulse waves, we overestimate the local Froude number at the free surface. Indeed, the amplitude of the wave has an influence on the propagation velocity of the leading waves. We thus define a modified local Froude number
\begin{equation}
\mathrm{Fr}_f^* = \frac{{v_f}_m}{\sqrt{g (h_0+A_m)}},
\end{equation}
as the ratio of the maximum velocity of the granular front ${v_f}_m$ and the solitary wave velocity $\sqrt{g(h_0+A_m)}$. Figure~\ref{Fig14}(a) shows the evolution of the relative maximum amplitude of the waves $A_m/h_0$ as a function of the modified local Froude number $\mathrm{Fr}_f^*$, where the linear scaling law
\begin{equation}
\frac{A_m}{h_0} \simeq 1.84~\mathrm{Fr}_f^*
\label{eq:modified-froude}
\end{equation}
captures well the experiments with a dispersion of about 10$\%$.

The ratio $\tau_c/\tau_w$ of the characteristic time of the collapse $\tau_c$ relative to the characteristic time of the wave generation $\tau_w$ is plotted as a function of the modified local Froude number $\mathrm{Fr}_f^*$ in figure~\ref{Fig14}(b). We see that this time ratio is always smaller than one, which suggests that the present experiments correspond to a subcritical regime where the generated wave propagates ahead of the moving granular collapse for all of the experiments presented in table 1. We also see that all data collapse fairly well on a straight line $\tau_c/\tau_w = \mathrm{Fr}_f^*$. This suggest that $\mathrm{Fr}_f^*$, defined as the ratio of the two typical velocities of the granular front and of the wave, corresponds also to the ratio of the two typical time scales for the evolution of the granular collapse and of the wave.

\subsection{A steady granular flow experiment}

Figures~\ref{Fig13} and~\ref{Fig14} highlight that the collective dynamics of the falling grains, in particular, the spreading dynamics of the granular front at the free surface, governs the generation of the leading wave. This result suggests that a continuous flow of grains entering the water, with a stationary granular front, would not generate any wave. To test this hypothesis, we have performed an experiment using the same set-up but now by opening the door only partially, on a small height $b=5$~cm compared with the initial height $H_0=60$~cm of the column, to establish a steady outflow of granular material like in the discharge from a silo. It has been shown that if the height of the grain column in the silo is larger than its width, the pressure at the bottom remains constant \citep{Bertho2003}, which thus ensures a constant discharge flow rate at a bottom outlet. For lateral openings at the bottom of rectangular containers, \citet{Zhou2017} have shown that the discharge flow obeys the usual Hagen-Beverloo law for $b/\ell < 1$ and is steady when $H(t)>b$. In our case, the opening $b=5$~cm is large enough compared with the grain diameter $d=5$~mm ($b/d=10$) to avoid any blockage or intermittent flow, and is thin enough when compared with the width of the channel $\ell=15$~cm ($b/\ell \simeq 0.3$) to recover the usual discharge law of Hagen--Beverloo during a large time enough, as $H_0\gg b$. The water height, $h_0=15$~cm, was also chosen large enough to prevent grains from accumulating too quickly at the bottom and reaching the opening.

\begin{figure}
  \centering
  \includegraphics[width=\hsize]{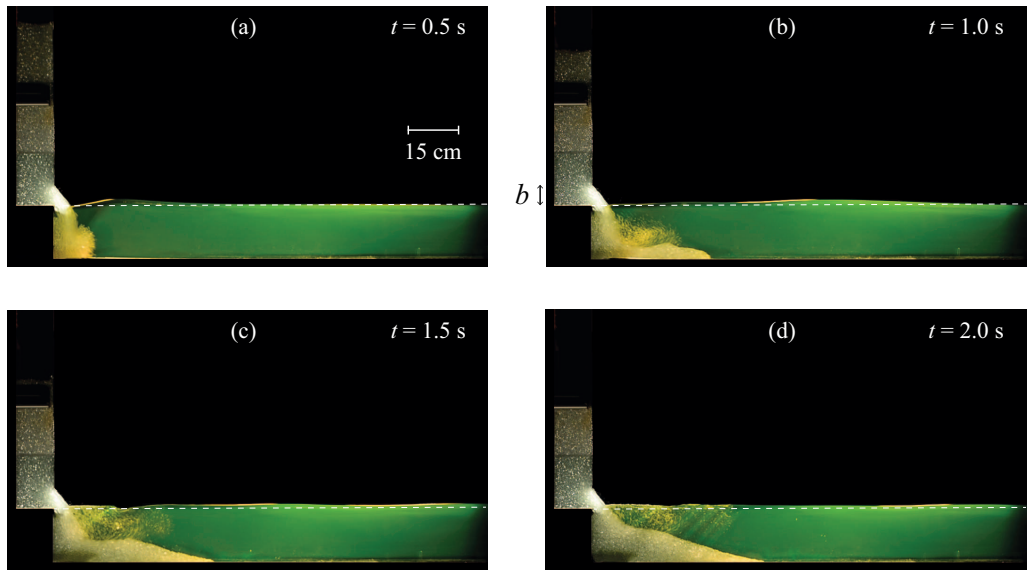}
  \caption{Image sequence of a continuous flow of grains entering into a water layer of height $h_0 = 15$ cm. The door is only opened at a small height $b=5$~cm. The white dashed line indicates the initial water free surface at rest.}
\label{Fig15}
\end{figure}

\begin{figure}
  \centering
  \includegraphics[width=\hsize]{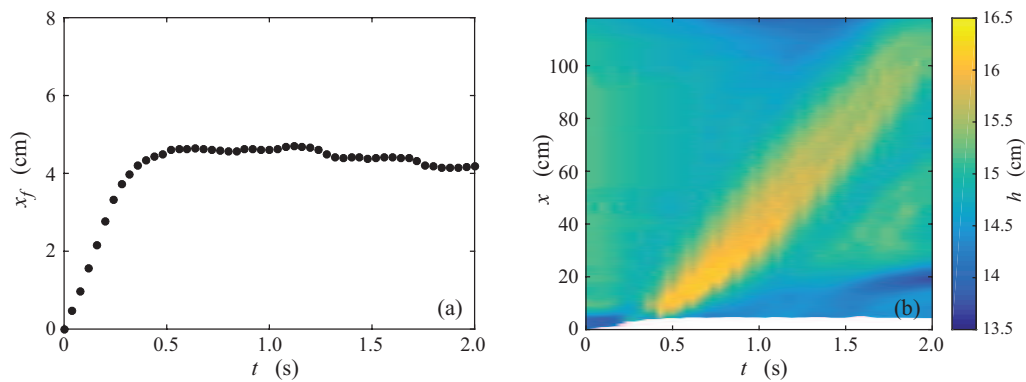}
  \caption{(a)~Time evolution of the front of the granular collapse $x_f$ and (b)~spatiotemporal evolution of the water free surface $h(x,t)$ during the continuous flow of grains into water. The white region corresponds to the spreading of the granular collapse and delimits the location $x_f$ of the granular front at $h_0=15$ cm.}
\label{Fig16}
\end{figure}

The experimental result of such a discharge flow is shown in figure~\ref{Fig15} where a time lapse illustrates the evolution of the granular discharge and the associated perturbation of the water free surface $h(x,t)$. In figure~\ref{Fig15}(a) we first observe the generation of a leading wave at the initial impact of granular material into water with a centimetric amplitude. Although the grains continue to flow into water, no significant secondary waves are then generated as shown in figures~\ref{Fig15}(b)--(d). Note that the experiment is stopped at $t \simeq 2$~s before all the grains have flowed into water. The evolution of the position of the granular front $x_f$ is plotted as a function of time in figure~\ref{Fig16}(a). We observe that beyond $t \simeq 0.5$~s the granular front becomes almost stationary, but in contrast with the case of figure~\ref{Fig11}, the velocity of the grains remains large. Figure~\ref{Fig16}(b) shows the spatiotemporal evolution of the water free surface $h(x,t)$ during the experiment. The maximum amplitude of the leading wave generated at the transient impact is $A_m \simeq 1.2$~cm, which is compatible with the scaling law of equation~(\ref{eq:modified-froude}). As previously emphasized in figure~\ref{Fig15}, no secondary wave is then generated from the continuous flow of grains into water once the granular front becomes stationary. Note that small fluctuations from the grain scale leads to very small ripples, which correspond to small capillary waves that are quickly damped by viscous dissipation.

This last experiment is in good agreement with the results emphasized in figures~\ref{Fig13} and \ref{Fig14}: the amplitude of the wave is governed by the velocity of the granular front at the free-surface. Therefore, a continuous flow of grains with a stationary front with a resulting zero velocity of the granular front only generates a leading wave at the initial transient, but no secondary waves as the grains continue to flow into water.

\section{Conclusion}

We have investigated experimentally the impulse generation of a tsunami-like solitary surface wave by the collapse of an initially dry column of grains into water of constant depth. The present results show that the collective entry of the granular material into water governs the wave generation. The collapse of the granular column under its own weight owing to gravity effects, with falling grains entering into water, makes a moving granular front that puts in motion the air/liquid interface in a similar way to a piston. The dynamics of this moving granular front depends on the initial geometry of the column, namely on its initial height but also on its aspect ratio. The results is that a scaling law is obtained between the rescaled maximal amplitude of the generated wave and a local Froude number based on the horizontal velocity of the granular front at the water surface. The scaling of the wave length appears more complex and would require further analysis. A first estimate of the energy transferred from the collapse of the initial granular column to the water wave reveals that the dissipation of energy arises mainly from the granular flow itself. The detailed mechanisms of the energy transfer from the grains to the water remains to be fully characterized. We highlight that the size and the density of the falling grains has a negligible influence on the wave amplitude, suggesting that the volume of grains entering into water is a key relevant parameter in the wave generation, especially in geophysical applications where the density of soils and rocks does not vary so much. A more detailed analysis of the influence of the volume entering into water in the wave generation may be used to constrain the amplitude of paleo-tsunamis or to predict the order of magnitude of a forthcoming event, as reported recently by \citet{RobbeSaule2020}.

We hope that the present experimental work will help improve numerical simulations of tsunami generated waves by landslides, that have been developed from some years, for instance, by \citet{Cremonesi2011}. These numerical simulations are essential to reproduce the complex natural topographies specific to each field location and to obtain reliable hazard assessment maps.\\

\textbf{Acknowledgments}

We are grateful to J.~Amarni, A.~Aubertin, L.~Auffray and R.~Pidoux for their contribution to the development of the experimental set-up, and P.~Claudin, P.~Aussillous, S.~Viroulet, and A.~Hildenbrand, for fruitful discussions. This work was supported by CNRS through its multidisciplinary program ``D\'efi Littoral'' via the projects ``SlideWave'' and ``SlideWave2''.\\

The authors report no conflict of interest.

\bibliographystyle{jfm}
\bibliography{biblioJFM}

\end{document}